\documentclass{aa}  

\usepackage{graphicx}
\usepackage{txfonts}
\begin{document} 

   \title{Optical observations of NEA 162173 (1999 JU3) during the 2011-2012 apparition\thanks{Photometric data is only available at the CDS via anonymous ftp to cdsarc.u-strasbg.fr (130.79.128.5) or via http://cdsarc.u-strasbg.fr/viz-bin/qcat?J/A+A/550/L11}}

   \author{Myung-Jin Kim \inst{1,2} \and
          Young-Jun Choi \inst{2}\thanks{corresponding author, \email{yjchoi@kasi.re.kr}} \and
          Hong-Kyu Moon \inst{2} \and
          Masateru Ishiguro \inst{3} \and
          Stefano Mottola \inst{4} \and
          Murat Kaplan \inst{5} \and
          Daisuke Kuroda \inst{6} \and
          Dhanraj S. Warjurkar \inst{3} \and
          Jun Takahashi \inst{7} \and
          Yong-Ik Byun \inst{1}
          }

   \institute{Department of Astronomy, Yonsei University, 50 Yonsei-ro, Seodaemun-gu, Seoul 120-749, Korea \\
              \email{skarma@galaxy.yonsei.ac.kr}\label{inst1} \and
              Korea Astronomy and Space Science Institute, 776 Daedeokdae-ro, Yuseong-gu, Daejeon 305-348, Korea \and
              Seoul National University, 1 Gwanak-ro, Gwanak-gu, Seoul 151-742, Korea \and
              German Aerospace Center (DLR), Rutherfordstra$\beta$e 2, 12489 Berlin, Germany \and
              Akdeniz Universitesi, Fen Fakultesi, Dumlupinar Bulvari, Kampus, 07058 Antalya, Turkey \and
              Okayama Astrophysical Observatory, National Astronomical Observatory of Japan, Asaguchi, Okayama 719-0232, Japan \and
              Nishi-Harima Astronomical Observatory, 407-2 Nishigaichi, Sayo-cho, Sayo-gun, Hyogo 679-5313, Japan
              }

   \date{Received 31 October 2012; Accepted  3 December 2012}


  \abstract
   {Near-Earth asteroid (hereafter NEA) 162173 (1999 JU3) is a potential target of two asteroid sample return missions, not only because of its accessibility but also because of the first C-type asteroid for exploration missions. The lightcurve-related physical properties of this object were investigated during the 2011 - 2012 apparition.}
   {We aim to confirm the physical parameters useful for JAXA's Hayabusa 2 mission, such as rotational period, absolute magnitude, and phase function. Our data complement previous studies that did not cover low phase angles.}
   {With optical imagers and 1-2 m class telescopes, we acquired the photometric data at different phase angles. We independently derived the rotational lightcurve and the phase curve of the asteroid.}
   {We have analyzed the lightcurve of 162173 (1999 JU3) (hereafter 1999 JU3), and derived a synodic rotational period of $7.625 \pm 0.003$ hr, the axis ratio $a/b = 1.12$. The absolute magnitude $H_\mathrm{R}$ = 18.69 $\pm$ 0.07 mag and the phase slope of $G = -0.09 \pm 0.03$ were also obtained based on the observations made during the 2011-2012 apparition. These physical properties are in good agreement with the previous results obtained during the 2007-2008 apparition.}
   {}

   \titlerunning{1999 JU3}
   \authorrunning{Myung-Jin Kim et al}
   \maketitle
 
%

\section{Introduction}
NEA 1999 JU3 is the primary target of JAXA's Hayabusa 2 mission and also a backup target of NASA's OSIRIS-REx mission. It is classified as a C-type asteroid (Binzel et al. 2001; Vilas 2008). Objects associated with this class have spectra similar to those of carbonaceous chondrite meteorites and are likely to contain abundant hydrous minerals and organics (Sato et al. 1997).

The results of lightcurve observations during the last opportunity in 2007 indicated that 1999 JU3 has a synodic rotational period of $7.6272 \pm 0.0072$ hr (Abe et al. 2008). From the mid-infrared observations, it was found that the asteroid has an effective diameter of $0.87 \pm 0.03$ km and a geometric albedo of $0.070 \pm 0.006$ (M\"{u}ller et al. 2011; Hasegawa et al. 2008). The pole orientation derived by the lightcurve observations still has a large uncertainty. Kawakami et al. (2010) obtained the pole axis of $\lambda=331\degr \pm 10\degr$ and $\beta=20\degr \pm 10\degr$ while M\"{u}ller et al. (2011) deduced $\lambda=73\degr$ and $\beta=-62\degr$, where $\lambda$ and $\beta$ are the ecliptic longitude and latitude of the pole orientation. Knowledge of these parameters is essential for the safety and fuel efficiency of landing and sampling on the asteroid. To confirm and update this information, it is critical to conduct observations of the asteroid during the successive apparitions. We had an observation opportunity for 1999 JU3 in the May to July 2012 season. It was the last chance to gather ground-based data before the launch of Hayabusa 2, which is scheduled in July 2014, with backup launch opportunities in December 2014 and in June and December 2015. 

In this paper, we outline our observations, data reduction, and analysis. We derived the rotational period and peak-to-peak variation from the lightcurve and the H-G parameters based on our data set. Because the previous lightcurve data is not publicly available at the moment, it is beyond the scope of this paper to update the value of pole orientation.
\begin{table*}
\caption{Observatory and instrument details}
\label{table:1}
\centering
\begin{tabular}{c c c c c c c}     
\hline\hline
Telescope\tablefootmark{a} & $\lambda$\tablefootmark{b} & $\phi$\tablefootmark{b} & Altitude & Instrument\tablefootmark{c} & Pixel scale & Observer\tablefootmark{d} \\
     &    &    & [m] & [CCD] & [$''$pix$^{-1}$] & \\
\hline
 UH 2.2 m & 204:31:40 & +19:49:34 & 4212.4 & Tek 2K   & 0.22  & MI \\
 CA 1.2 m & 2:32:45   & +37:13:25 & 2173.1 & e2v 4K   & 0.63  & SM \\
 NH 2 m   & 134:20:08 & +35:01:31 & 435.9  & e2v 2K & 0.32  & JT, MI \\
 TUG 1 m  & 30:19:59  & +36:49:31 & 2538.6 & SI 4K    & 0.62  & MK, MJK, SK, OU, EG \\
 HCT 2 m  & 78:57:51  & +32:46:46 & 4500.0 & SITe 2K  & 0.17  & DSW, MI \\
\hline
\end{tabular}
\tablefoot{
\tablefoottext{a}{Abbreviations: UH = University of Hawaii, CA = Calar Alto,  NH = Nish-Harima, TUG = Tubitak Ulusal Gozlemevi (Turkish National Observatory), HCT = Himalayan Chandra Telescope} 
\tablefoottext{b}{Eastern longitude and geocentric latitude of each observatory} 
\tablefoottext{c}{Both e2v 4K CCD and SI 4K CCD were configured with 2 $\times$ 2 binning} 
\tablefoottext{d}{Observer: DSW = D. S. Warjurkar, EG = G. Guzel, JT = J. Takahashi, MI = M. Ishiguro, MK = M. Kaplan, MJK = M.-J. Kim, OU = O. Uysal, SK = S. Kaynar, SM = S. Mottola}
}
\end{table*}
\begin{table*}
\caption{Observational circumstances}
\label{table:2}
\centering
\begin{tabular}{c c c c c c c c c c c c}     
\hline\hline
UT date & RA & DEC & L$_\mathrm{PAB}$ & B$_\mathrm{PAB}$ & $\alpha$ & r & $\Delta$ & V & Telescope & Seeing & Sky \\
(DD/MM/YY) & [hr] & [\degr] & $ [\degr] $ & [\degr] & [\degr] &  [AU]  & [AU] & [Mag] &   & [\arcsec] & condition \\
\hline
 06.9/08/2011 & 2.58 & +25.62 & 196.9 & 7.9 & 54.7 & 1.242 & 0.730 & 20.96 & UH 2.2 m & 1.3 & Photometric \\
 06.9/06/2012 & 16.42 & -20.64 & 251.1 & 0.6 & 6.3 &  1.376 & 0.364 & 18.16 & CA 1.2 m & 2.2 & Clear \\
 08.9/06/2012 & 16.35 & -20.06 & 251.1 & 0.7 & 8.5 &  1.379 & 0.369 & 18.29 & CA 1.2 m & 3.2 & Clear \\
 20.9/06/2012 & 16.03 & -17.07 & 251.9 & 2.3 & 20.2 & 1.394 & 0.413 & 18.98 & CA 1.2 m & 2.6 & Clear \\
 21.9/06/2012 & 16.01 & -16.89 & 252.0 & 2.0 & 20.9 & 1.395 & 0.417 & 19.03 & TUG 1 m & 2.8 & Clear \\
 21.9/06/2012 & 16.01 & -16.87 & 252.0 & 2.4 & 21.0 & 1.395 & 0.417 & 19.03 & CA 1.2 m & 2.0 & Clear \\
 22.5/06/2012 & 16.00 & -16.75 & 252.2 & 2.4 & 21.5 & 1.395 & 0.420 & 19.06 & NH 2 m & 3.1 & Cirrus \\
 22.9/06/2012 & 15.99 & -16.67 & 252.2 & 2.5 & 21.8 & 1.396 & 0.422 & 19.08 & CA 1.2 m & 2.3 & Clear \\
 24.9/06/2012 & 15.96 & -16.30 & 252.4 & 2.7 & 23.4 & 1.398 & 0.432 & 19.18 & CA 1.2 m & 2.4 & Clear \\
 04.8/07/2012 & 15.88 & -14.92 & 252.3 & 3.5 & 30.2 & 1.406 & 0.485 & 19.65 & TUG 1 m & 2.5 & Clear \\
 17.7/07/2012 & 15.91 & -14.09 & 257.9 & 4.2 & 36.6 & 1.413 & 0.566 & 20.18 & HCT 2 m & 1.2 & Photometric \\
 18.7/07/2012 & 15.92 & -14.07 & 258.2 & 4.3 & 37.0 & 1.414 & 0.573 & 20.21 & HCT 2 m & 1.2 & Cirrus \\
 19.7/07/2012 & 15.93 & -14.05 & 258.5 & 4.4 & 37.3 & 1.414 & 0.579 & 20.25 & HCT 2 m & 1.3 & Cirrus \\
\hline
\end{tabular}
\tablefoot{$UT$ date corresponding to the mid time of the observation, J2000 coordinates of 1999 JU3 ($RA$ and $DEC$), Phase Angle Bisector (PAB) - the bisected arc between the Earth-asteroid and Sun-asteroid lines - ecliptic longitude ($L_\mathrm{PAB}$) and ecliptic latitude ($B_\mathrm{PAB}$), the solar phase angle ($\alpha$), the helicentric ($r$) and the topocentric distances ($\Delta$), the apparent predicted magnitude ($V$), average seeing and sky condition.}
\end{table*}

\section{Observations}
Observations of 1999 JU3 were carried out for 13 nights during its apparition in 2011 and 2012, with 1-2 m class telescopes using CCD cameras with the University of Hawaii (UH) 2.2 m telescope in Hawaii, USA, the Calar Alto (CA) Astronomical Observatory 1.2 m telescope in Almeria, Spain, the Nish-Harima (NH) Astronomical Observatory Nayuta 2 m telescope in Sayo, Japan, the Tubitak Ulusal Gozlemevi (TUG) 1 m telescope in Bakirlitepe, Turkey, and the 2 m Himalayan Chandra Telescope (HCT) in Hanle, India. 
The details of the observatories including instruments are shown in Table 1. 
Out of 13 nights of our observations, the UH data were obtained in 2011, while the other observations just after opposition were mostly conducted during the June to July period in 2012. 
Because 1999 JU3 was relatively faint in 2011, all images taken with the UH 2.2 m telescope were acquired through the nonsidereal tracking mode that corresponds to the predicted motion of the object. 
The CA 1.2 m telescope was tracked with a tracking vector halfway between sidereal rate and that of 1999 JU3. 
In this way both 1999 JU3 and the background stars were trailed by the same amount.
The three other telescopes were guided at sidereal rates, and the exposure time was determined by two factors: the apparent motion of the asteroid was set to be less than the FWHM of the stellar profiles at each observatory; the signal-to-noise ratio of the object was adjusted to be greater than 30.
Accordingly, during the observations made at the sidereal rate, the maximum exposure never exceeded 300 seconds. 

The details of observations are listed in Table 2: the date ($UT$), the median coordinates ($RA$ and $DEC$, J2000), phase angle bisector (PAB), ecliptic longitude ($L_\mathrm{PAB}$), ecliptic latitude ($B$$_\mathrm{PAB}$), the solar phase (sun-asteroid-observer's) angles ($\alpha$), heliocentric- ($r$) and topocentric- ($\Delta$) distances, the apparent predicted magnitudes ($V$), the average size of seeing disk, and the sky condition. Conducting observations in a wide range of solar phase angles is of utmost importance when calculating both the absolute magnitude ($H$) and the slope parameter ($G$). 
For this purpose, we obtained data of 1999 JU3 at different phase angles, between 6 and 54 degrees. 
The weather during the observational runs was mostly clear, however, on the nights of 22 June 2012 in Japan and 18-19 July 2012 in India, we had cirrus. 
Photometric standard stars were observed on 17 July 2012 with the HCT 2 m telescope and 6 August 2011 with the UH 2.2 m telescope.
We made time-series observations with the Johnson R filter in order to characterize the rotational status of the asteroid because the combinations of R-band and optical imagers provide the highest sensitivity for asteroidal surfaces. 
During the observational windows in the June-July 2012 season, we were not able to cover the whole rotational period on a single night in the northern hemisphere due to shorter visibility. 

During the 20-24 June period, we organized three observatories in Japan, Turkey, and Spain in order to perform consecutive observations to cover the entire lightcurve. 
In addition, the five-day continuous observation conducted at CA enabled us to obtain a full-phased lightcurve with a sufficient signal-to-noise ratio. 
Observations dedicated to calibration for different telescopes were carried out in Turkey on August 2012. 
To calibrate all the data gathered from various observatories, the same CCD fields imaged in June and July 2012 were taken on a single photometric system at the TUG 1 m telescope.

\section{Data reduction}
All the data reduction procedures were performed using the Image Reduction and Analysis Facility (IRAF) software package. 
Bias and dark frames with relatively high-standard deviations were excluded in our analysis. 
Twilight sky flats were derived before sunrise and after sunset, and combined to create a master flat image for each night. 
The lightcurves of 1999 JU3 were constructed based on relative magnitude, which is the difference between instrumental magnitude of the asteroid and an average magnitude of each comparison star.
To choose a set of comparison stars, we inspected whether there is any short-term variability in the brightness of a star, according to the following method. 
We overlapped each frame that was taken during the same night in order to calculate differences in magnitude of each star and the median value of the ensemble of the remaining stars. 
In such a way, we were able to determine standard deviations in each frame, and hence checked if they had variability during the course of single night's observation. 
Likewise, we repeatedly applied the same procedure to the other CCD fields, and successfully found stars with minimum standard deviation. We may thus conclude that they have the smallest amount of light variation in each field. 
Finally, we selected three to five comparison stars with the typical scatter in the comparison star's magnitudes of 0.01 - 0.02. 

The resulting lightcurve was reconstructed based on the magnitudes of the comparison stars computed from the calibration images taken on the 13 August 2012. 
Then, we computed the offsets that occurred by magnitude corrections in each frame and applied them to all fields.
In the next step, differential offsets were applied to the combined lightcurve. 
The final step in our data reduction process is observation time ($UT$) correction for light travel time.
Standard star observations of the Landolt (1992) catalog stars were obtained with the HCT 2 m telescope on 17 July 2012 and with the UH 2.2 m telescope on 6 August 2011 in photometric conditions. 
The photometry of standard stars was performed with an aperture of 12 arcsec. Using the IRAF photcal package, the zeropoints, atmospheric extinction coefficients, and color terms for Johnson $B$, $V$, and $R$ filters were calculated for each night.

\section{Analysis and results}
\subsection{Rotational period and lightcurve}
To find the periodicity in the lightcurve, the fast chi-squared ($F_\mathrm{\chi^2}$) technique (Palmer 2009) was employed. 
In addition, we checked the result with the discrete Fourier transform algorithm (Lenz and Breger 2005). 
These two different techniques produced similar results, e.g., a rotational period of 7.6248 hr and 7.6251 hr, respectively, whose discrepancy is only one second. 
A single strong peak found at $PF_\mathrm{\chi^2}$ = 3.14763 cycle/day = 7.6248 hr seems to be consistent with 0.3178 days from the previous observation conducted during the apparition in the 2007 - 2008 season (Abe et al. 2008).
We also obtained the highest spectral power at 6.29498 cycle/day with the discrete Fourier transform algorithm, which corresponds to a rotational period of 7.6251 hr assuming a double-peaked lightcurve.

Before combining the lightcurves acquired at different solar phase angles, we investigated the effect of changing the viewing longitude (Magnusson et al. 1989).
However, considering the geometry of the Sun, the Earth, and the asteroid  in June - July 2012 (see Table 2), the maximum difference between the observed synodic period and the sidereal period (Harris et al. 1984) was found to be  0.0013 hr, which is negligible and within the statistical errors.
We present the resulting composite lightcurve of 1999 JU3 in Fig. 1, which is folded with the period of $7.625 \pm 0.003$ hr at the epoch of JD 2456106.834045. 
We combined the data obtained from different telescopes and instruments and computed relative magnitudes based on our observations of comparison stars. We also made use of the observation conducted on 12 June 2012 as a reference for calibration procedures.

The peak-to-peak variations in magnitude are caused by change in the apparent cross-section of the rotating tri-axial ellipsoid, with semi-axes $a$, $b$, and $c$, where $a > b > c$ when the body rotates along the $c$ axis. 
The lower limit of axis ratio $a/b$ can be expressed as $a/b =10^{0.4\Delta m}$ (Binzel et al. 1989).
The peak-to-peak variation in the lightcurve becomes larger with the increase in the solar phase angle. 
For this reason, we divided our lightcurve data into three groups: 1) early June around 7 deg phase, 2) end of June around 21 deg phase, and 3) mid July around 37 deg phase.
The following empirical relationship between those two parameters was found by Zappala et al. (1990).
\begin{equation}
A(0\degr)=\frac{A(\alpha)}{(1+m\alpha)}
\end{equation}
where $A(0\degr)$ is the amplitude of lightcurve at zero phase angle and $A(\alpha)$ is the amplitude measured at solar phase angle = $\alpha$. 
Zappala et al. (1990) found best-fit $m$-values of 0.030, 0.015, and 0.013 degree$^\mathrm{-1}$ for S, C, and M-type asteroids, respectively. 
We obtained the same result of the amplitude $A(0\degr)=0.12$ from those three groups with the $m$-value of 0.015 for an average C-type asteroid, which implies a lower bound for the $a/b$ axis ratio of 1.12.

   \begin{figure}
   \centering
   \includegraphics[width=\hsize]{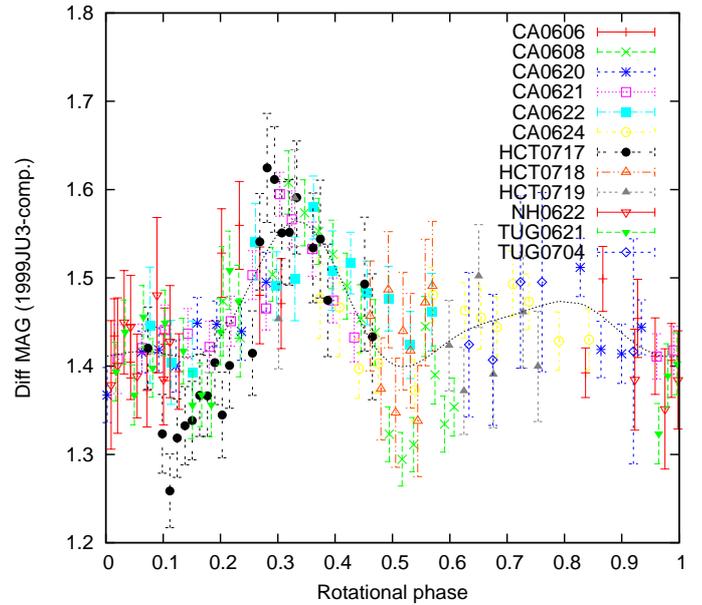}
      \caption{Composite lightcurve of 1999 JU3 folded at the period of $7.625 \pm 0.003$ hr at the zero epoch of JD 2456106.834045 obtained from the 2012 apparition. The dotted line is a fit to the fourth-order Fourier model using the fast chi-squared technique.
              }
         \label{FigVibStab}
   \end{figure}

   \begin{figure}
   \centering
   \includegraphics[width=\hsize]{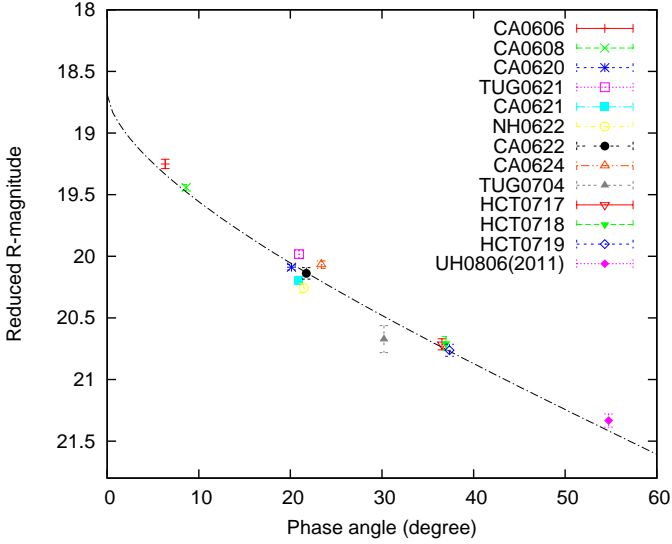}
      \caption{Phase function of 1999 JU3. Each data point represents the reduced R-band magnitudes at corresponding phase angles. The long dashed dotted line represents the IAU ($H, G$) phase function fit, where $H_\mathrm{R} = 18.69 \pm 0.07$ mag and $G = -0.09 \pm 0.03$.
              }
         \label{FigVibStab}
   \end{figure}

\subsection{$H-G$ parameters and size}
To reproduce the phase curve of 1999 JU3, all the data were calibrated on the basis of absolute magnitudes of comparison stars obtained from the photometric observation either on 17 July 2012 with the HCT 2 m telescope or on 6 August 2011 with the UH 2.2 m telescope. 
Our observations covered the phase angle from 6\degr to 55\degr . 
We note that the previous works did not contain the data at low phase angles ($\alpha < 20 \degr $). 
Figure 2 shows the phase angle dependence of the reduced magnitude of 1999 JU3. 
Owing to the narrow visibility window in the northern hemisphere during the 2012 apparition, our data obtained on a single night did not encompass a full rotational phase. 
Therefore we plotted the zeroth-order term for each lightcurve, which provides a good approximation of the mean value from the Fourier fit of the composite lightcurve.
Then, we calculated the best fit parameters of $H_\mathrm{R}=18.69 \pm 0.07$ mag and $G=-0.09 \pm 0.03$ with a linear least squares fit.
From the observation made during the 2007 - 2008 apparition, Abe et al. (2008) also found similar values of $H=18.82 \pm 0.02$ mag and $G=-0.11 \pm 0.01$.
The slope parameter of a typical C-type asteroid is $G=0.04 \pm 0.06$ (Harris et al. 1989; Lagerkvist and Magnusson, 1990); however, there is still a large uncertainty in determining the G value.
The diameter $D$ of an asteroid can be calculated through the relation between absolute magnitude and visual geometric albedo (Bowell and Lumme, 1979):

\begin{equation}
D_\mathrm{(km)}=\frac{1329}{\sqrt{p_\mathrm{v}}} 10^{-0.2H_\mathrm{v}}
\end{equation}

In practice, the size of an asteroid is more precisely determined by thermal IR observations. 
If we adopt the size of 1999 JU3 as $0.87 \pm 0.03$ km (M\"{u}ller et al. 2011), the R-band geometric albedo of $0.078 \pm 0.013$ can be obtained using Eq. (2).

\section{Conclusions}

We have introduced the composite lightcurve of 1999 JU3 and found a synodic rotational period of $7.625 \pm 0.003$ hr. 
Since the longitude and latitude of the PAB changed little during the period covered by our 2012 observations (about 10 deg and 4 deg, respectively), we used the whole 2012 data set for our analysis.
When we compared the period of $7.6272 \pm 0.0072$ hr derived from the previous study (Abe et al. 2008) with ours, it was found that the precision has been slightly increased.

In addition, we re-analyzed the 1999 JU3 data obtained in the Subaru-Mitaka-Okayama-Kiso Archive System (SMOKA) (Baba et al. 2002) during the September - November season in 2007, with the same data reduction method. 
To compare the lightcuves obtained from both apparitions, we overlaid one on the other, with the same best-fit period of 7.6275 hr at the epoch of JD 2454348.248074 for 2007 and 2456085.41352 for 2012.
Overall shapes of individual lightcurves look very similar to each other (See Figure 3).
It might be regarded as a clue that we were looking at the asteroid from similar aspect angles during each apparition.

Based on our new data, we presented the phase curve of 1999 JU3 encompassing a wide range of viewing geometry, together with the absolute magnitude $H_\mathrm{R} = 18.69 \pm 0.07$ mag and the slope parameter $G = -0.09 \pm 0.03$. 

   \begin{figure}
   \centering
   \includegraphics[width=\hsize]{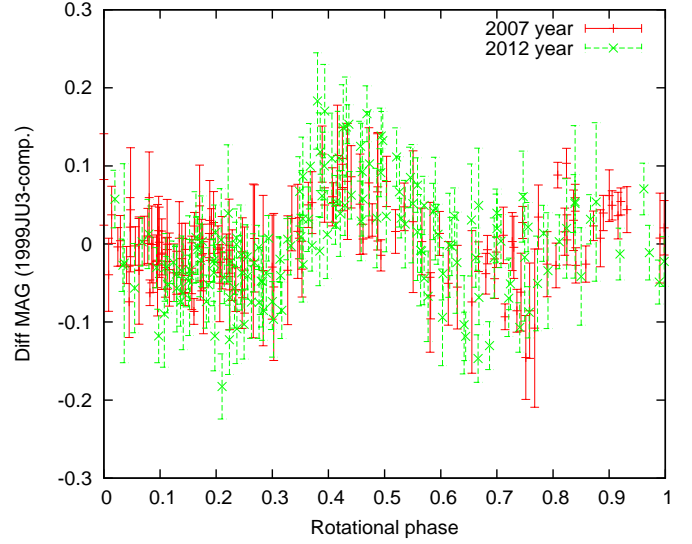}
      \caption{Superposed lightcurve based on the data from 2007 and 2012, folded with the period of 7.6275 hr at the epochs of JD 2454348.248074 and 2456085.41352, respectively.
              }
         \label{FigVibStab}
   \end{figure}

\begin{acknowledgements}
      MJK was partly supported by Korea Research Council of Fundamental Science $\&$ Technology. YJC was partly supported by the collaborative research project between NRF-JSPS (2012010270). YIB acknowledges the support of NRF grant 2011-0030875. This study was based [in part] on data collected at [Subaru Telescope|Okayama Astrophysical Observatory|Kiso observatory (University of Tokyo)] and obtained from the SMOKA, which is operated by the Astronomy Data Center, National Astronomical Observatory of Japan. 
\end{acknowledgements}

\end{document}